\documentclass[aps,prl,10pt,twocolumn,superscriptaddress,preprintnumbers,footinbib]{revtex4-1}

\usepackage[squaren]{SIunits}
\usepackage{graphicx,SIunits}
\usepackage{dcolumn}
\usepackage{bm}
\usepackage{amssymb,amsmath,latexsym}
\usepackage{float}

\newcommand{\tn}{\textnormal}

\begin{document}

\title{Observation of Generalized Optomechanical Coupling and Cooling on Cavity Resonance}

\author{Andreas Sawadsky}
\affiliation{Institut f\"{u}r Gravitationsphysik, Leibniz Universit\"{a}t Hannover and Max-Planck Institut f\"{u}r Gravitationsphysik (Albert-Einstein Institut),\\ Callinstra\ss{}e 38, D-30167 Hannover, Germany}

\author{Henning Kaufer}
\affiliation{Institut f\"{u}r Gravitationsphysik, Leibniz Universit\"{a}t Hannover and Max-Planck Institut f\"{u}r Gravitationsphysik (Albert-Einstein Institut),\\ Callinstra\ss{}e 38, D-30167 Hannover, Germany}

\author{Ramon Moghadas Nia}
\affiliation{Institut f\"{u}r Gravitationsphysik, Leibniz Universit\"{a}t Hannover and Max-Planck Institut f\"{u}r Gravitationsphysik (Albert-Einstein Institut),\\ Callinstra\ss{}e 38, D-30167 Hannover, Germany}\affiliation{Vienna Center for Quantum Science and Technology (VCQ), Faculty of Physics, University of Vienna, Boltzmanngasse 5, 1090 Vienna, Austria\\}

\author{Sergey P. Tarabrin}
\affiliation{Institut f\"{u}r Gravitationsphysik, Leibniz Universit\"{a}t Hannover and Max-Planck Institut f\"{u}r Gravitationsphysik (Albert-Einstein Institut),\\ Callinstra\ss{}e 38, D-30167 Hannover, Germany}
\affiliation{Institut f\"ur Theoretische Physik, Leibniz Universit\"at Hannover, Appelstra\ss{}e 2, D-30167 Hannover, Germany}

\author{Farid Ya.~Khalili}
\affiliation{Moscow State University, Department of Physics, Moscow 119992, Russia}

\author{Klemens Hammerer}
\affiliation{Institut f\"{u}r Gravitationsphysik, Leibniz Universit\"{a}t Hannover and Max-Planck Institut f\"{u}r Gravitationsphysik (Albert-Einstein Institut),\\ Callinstra\ss{}e 38, D-30167 Hannover, Germany}
\affiliation{Institut f\"ur Theoretische Physik, Leibniz Universit\"at Hannover, Appelstra\ss{}e 2, D-30167 Hannover, Germany}

\author{Roman Schnabel}
\email[]{Corresponding author: roman.schnabel@aei.mpg.de}
\affiliation{Institut f\"{u}r Gravitationsphysik, Leibniz Universit\"{a}t Hannover and Max-Planck Institut f\"{u}r Gravitationsphysik (Albert-Einstein Institut),\\ Callinstra\ss{}e 38, D-30167 Hannover, Germany}




\date{\today}

\begin{abstract}
Optomechanical coupling between a light field and the motion of a cavity mirror via radiation pressure plays an important role for the exploration of macroscopic quantum physics and for the detection of gravitational waves (GWs). It has been used to cool mechanical oscillators into their quantum ground states and has been considered to boost the sensitivity of GW detectors, e.g. via the optical spring effect. Here, we present the experimental characterization of generalized, that is,  dispersive and dissipative, optomechanical coupling, with a macroscopic (1.5\,mm)$^2$-size silicon nitride membrane in a cavity-enhanced Michelson-type interferometer. We report for the first time strong optomechanical cooling based on dissipative coupling, even on cavity resonance, in excellent agreement with theory. Our result will allow for new experimental regimes in macroscopic quantum physics and GW detection.
\end{abstract}

\pacs{}

\maketitle


Optomechanical cavities \cite{Aspelmeyer2014, Chen2013, Meystre2013}, whose mirrors are explicitly able to move, have been suggested to improve the sensitivity of gravitational wave detectors beyond the free-mass standard quantum limit (SQL) \cite{Braginsky1967,Braginsky1970,Braginsky1997,Buonanno2002},  
to test modified models of quantum mechanics \cite{Bahrami2014,Bateman2013,Marshall2003,Nimmrichter2011a,Nimmrichter2014,Romero-Isart2011b}, and to realize applications in quantum information processing \cite{Aspelmeyer2014,Rips2013,Stannigel2010,Stannigel2011}.
In such cavities, the motion of the mirror dynamically changes the cavity parameters and thus the power of the cavity field. The power change, in turn, couples back to the motion of the mirror, thereby creating a (dynamical) optomechanical coupling of the optical and mechanical degrees of freedom.
Two mechanisms can be distinguished. First, the displacement of the mirror changes the resonance frequency of the cavity, leading to so-called dispersive coupling \cite{Aspelmeyer2014}. Second, the displacement of the mirror changes the linewidth of the cavity, leading to so-called dissipative coupling \cite{Elste2009}.
Up until now, optomechanics was mainly investigated in the limit of strongly dominant dispersive coupling. This regime, however, shows significant constraints.
First, optical ground-state cooling \cite{Chan2011}, so far, is based on dispersive coupling, and thus requires a red-detuned light field and a cavity whose linewidth is smaller than the mechanical frequency (sideband-resolved regime) \cite{Marquardt2007,Wilson-Rae2007}. These requirements are unfeasible  for low mechanical frequencies, i.e.~in the interesting regime of macroscopic and heavy oscillators. Second, designs of next-generation gravitational wave (GW) detectors with sensitivities enhanced by the optical spring \cite{Harry2010,Accadia2011, Somiya2011} consider only dispersive coupling so far \cite{Buonanno2002a}. The optical spring, however, was found to be inherently unstable \cite{Buonanno2002}, which results in uncontrolled motions of the pendulum-suspended mirrors, and requires a yet-to-be-developed control system in order to exploit optomechanical effects for achieving sensitivities beyond the SQL.
Generalized optomechanical systems with significant contributions from both dispersive and dissipative couplings significantly broaden the scope of optomechanics. In such systems, strong optical cooling on cavity resonance is predicted, making the sideband-resolved regime \cite{Elste2009,Xuereb2011,Weiss2013a,Tarabrin2013} unnecessary. In Refs. \cite{Tarabrin2013,Vostrosablin2014} it was shown that the interference of dispersive and dissipative coupling when operating close to the dark fringe can produce a stable optical spring in GW detectors. Such a setup would improve the sensitivity beyond the SQL without the need for a control system or additional light beams \cite{Rehbein2008}.
The application of generalized optomechanical systems is thus wide-ranging and an experimental test of its mathematical description is essential.
Recently, dispersive and dissipative couplings were observed with nanomechanical oscillators \cite{Li2009,Wu2014}. However, in these experiments the dissipative coupling was dominated by internal dissipation due to photon loss into unaccessible channels. Unique features such as optical cooling on resonance were not observed.

In this Letter we report on the experimental characterization of generalized optomechanical coupling in a macroscopic system of high relevance for GW detection and macroscopic quantum physics. We use a Michelson-Sagnac interferometer (MSI) with a detuned signal-recycling cavity to vary the weighting of dispersive and dissipative coupling between the light field and a silicon nitride (SiN) membrane, and compare our experimental data with the theoretical model. In contrast to previous works dissipative coupling in our setup is not due to internal dissipation, and in principle all photons are detectable in the output ports.
For the first time, we observe optical cooling on cavity resonance providing evidence for the possibility of achieving optical cooling of massive low frequency oscillators, as well as a stable optical spring.

\begin{figure}[h]
\includegraphics [scale=1.05] {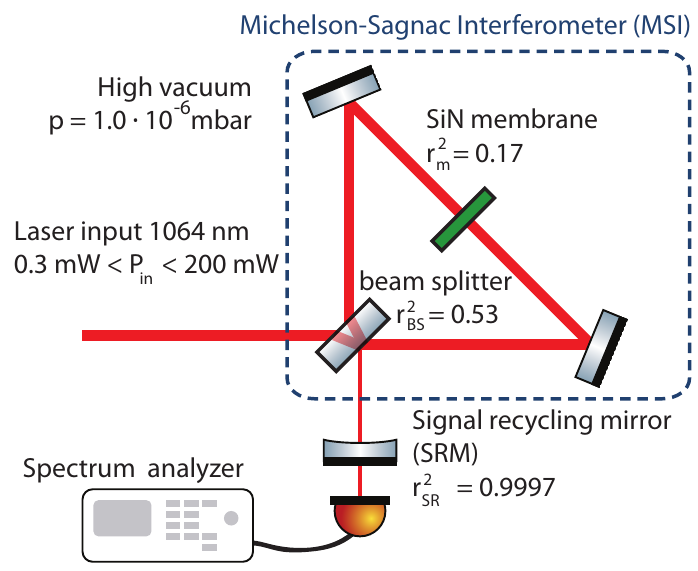}
\caption{Schematic of the optomechanical setup. Laser light is split into two beams which are directed towards a translucent and partially retroreflecting membrane. Altogether four light beams, which are either reflected or transmitted through the membrane, interfere at the beam splitter, thereby forming a Michelson-Sagnac interferometer. The interferometer corresponds to a compound mirror whose effective reflectivity depends on the position of the membrane. Together with a signal-recycling mirror (SRM) in the interferometer's output port, the complete setup allows for tuning from strong dispersive to strong dissipative optomechanical coupling.}
\label{setup}
\end{figure}

Our optomechanical setup is shown in Fig.\ref{setup}. It represents a Michelson-Sagnac interferometer that contains a SiN membrane as a movable translucent mirror with a surface of $1.5\times \unit{1.5}{\milli\metre}^2$.
The optical and optomechanical properties of this interferometer type were presented in \cite{Westphal2012, Friedrich2011, Kaufer2012, Kaufer2013}. In contrast to our previous realizations of the Michelson-Sagnac topology, the present work uses an unbalanced beam splitter with a reflectivity of $r^2_{\tn{BS}}=0.53$.
According to theory, the unbalanced splitting is expected to increase the influence of the dissipative optomechanical coupling \cite{Tarabrin2013}, which facilitates reaching the regime of generalized optomechanics.
The movable mirror is a SiN membrane, which has a power reflectivity of $r^2_{\tn{m}} = 0.17$ at normal incidence at the laser wavelength of $\lambda_{\tn{0}}=\unit{1064}{\nano\metre}$. Its subwavelength thickness of $\unit{40}{\nano\metre}$ results in an effective mass of $m_{\tn{eff}}=\unit{80}{\nano\gram}$ for the fundamental frequency of oscillation $f_{\tn{m}}=\unit{136}{\kilo\hertz}$.
The membrane served as a common endmirror for both arms of the Michelson mode. Transmitted light excites the Sagnac mode.
The membrane position influences the interference condition of Michelson and Sagnac modes in the interferometer output port.
Assuming a perfect interferometer contrast and taking into account the reflectivity of the membrane, the normalized transmitted power $t^2_{\tn{MSI}}=P_{\tn{out}}/P_{\tn{in}}$ ranges from $0$ to $0.17$ \cite{Friedrich2011}.
The Michelson-Sagnac interferometer thus represents a compound mirror whose reflectivity ranges from 83\% to 100\% (assuming perfect visibility), depending on the position of the membrane.

\begin{figure}[h]
\includegraphics [scale=0.7] {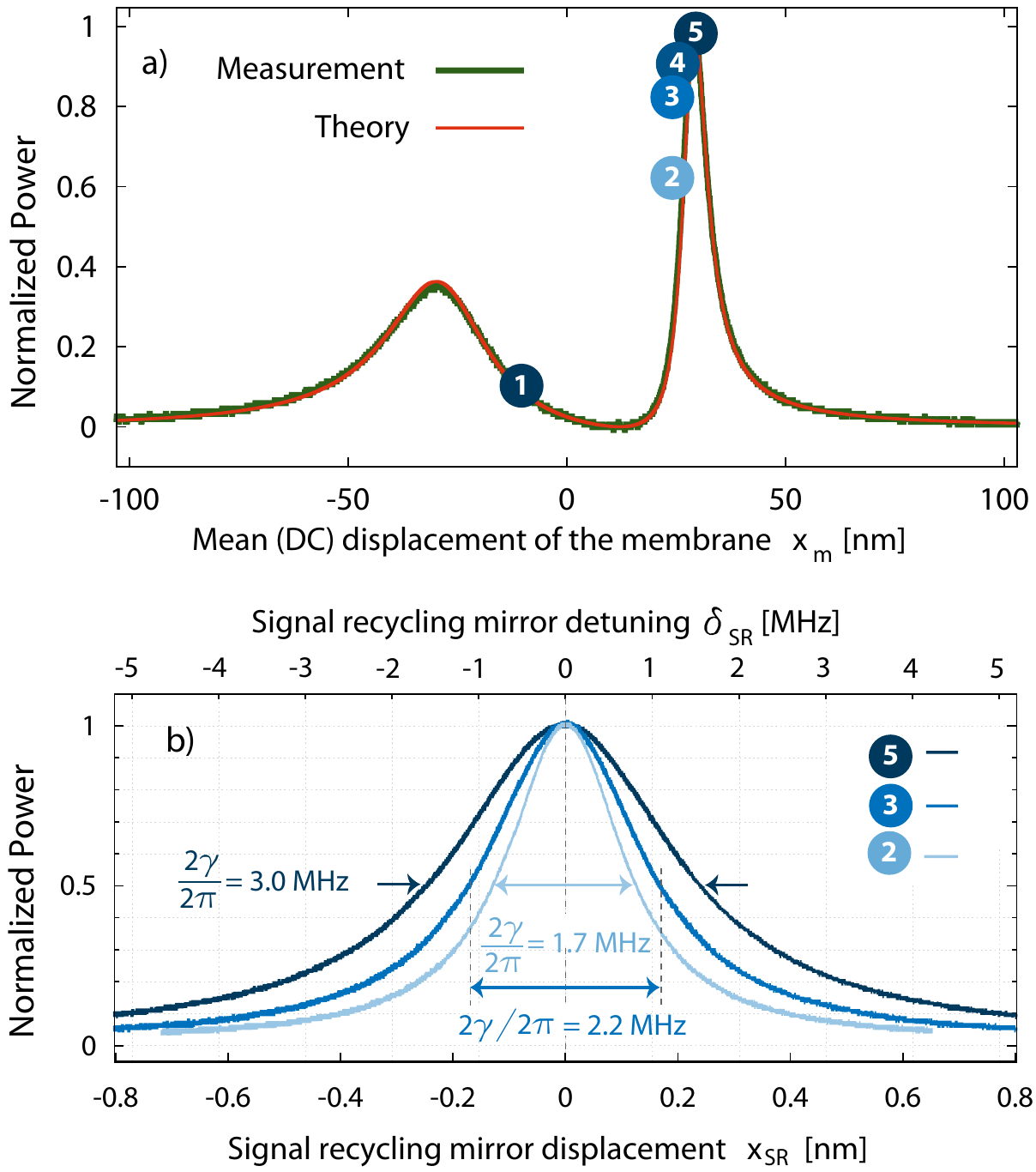}
\caption{Measured normalized intracavity (transmitted) light powers versus mean membrane displacement (a) and versus signal recycling mirror displacement for three membrane displacements (b). The input powers used for these measurements were \unit{20} (a) and \unit{5}{\milli\watt} (b), respectively. In (a) the SRM was positioned such that at operation point $5$ an impedanced matched resonator was achieved. Numbers $1$ to $5$ mark operating points of the membrane that we used for the measurements in this work. Point $1$ is at $7\%$, $2$ at $60\%$, $3$ at $90\%$, $4$ at $95\%$ and $5$ at $100\%$ of the cavity resonance peak height. Panel (b) shows the dependence of the cavity linewidth $\gamma$ on the three different chosen membrane positions $2$, $3$ and $5$ exemplary.
}
\label{outpower}
\end{figure}

In the present work we  combined the Michelson-Sagnac interferometer with an additional mirror ($r^2_{\tn{SR}}=0.9997$), thereby forming an optical cavity. The concept is adopted from "signal-recycling"  in gravitational wave detectors \cite{Meers1988}.
For the cavity-enhanced Michelson-Sagnac interferometer the transmitted power now depends on \emph{two} tunable parameters:  the position of the membrane and the position of the signal-recycling mirror (SRM).  Its behavior is shown in Fig.\ref{outpower}. A shift of the position of the SRM changes the cavity resonance frequency. The displacement of the membrane also changes the cavity resonance frequency, but in particular it too changes the cavity linewidth $\gamma$ (half-width at half maximum) \cite{supplementary}. Accordingly, the detuning between the cavity and laser frequency can be changed by displacing of either the SRM or the membrane. At the membrane positions $1$ to $5$ the linewidth  $\gamma/2\pi$ is  tunable from \unit{0.7 - 1.5}{\mega\hertz}. The cavity enhanced setup was thus far from the sideband-resolved regime, which requires $\gamma\ll 2\pi f_{\tn{m}}$,  and which is necessary to reach the mechanical ground state in experiments operating in the limit of purely dispersive optomechanical coupling. In Fig.\ref{outpower} (b) we illustrate the linewidth measurement for membrane positions $2$, $3$ and $5$, exemplary. For these measurements we put the membrane to the given positions and after that we scanned the position of the signal recycling mirror with a piezoactuator. To calibrate the x axis in Fig.\ref{outpower} (b) into frequency we measured for the same membrane positions the transfer functions of the cavity with a spectrum analyzer. Together with the tunable reflectivity of the interferometer we reached a maximal finesse of about $1200$. This value was not limited by the contrast when the membrane was set to maximal destructive interference at the output port but by losses in optical components ($0.5\,\%$ in total). The effective cavity length resulted in a free-spectral range of $\tn{FSR}=c/(2\mathcal{L})=\unit{1.7}{\giga\hertz}$ with an effective cavity length of $2\mathcal{L}=\unit{0.174}{\meter}$.
The complete interferometer according to Fig.~\ref{setup}  was set up in a high-vacuum environment ($p=\unit{1.0\cdot 10^{-6}}{\milli\bbar}$) to avoid damping of the oscillator motion by residual gas.
In the absence of the SRM (i.e.~with neither optical cooling nor heating) the oscillator mechanical quality factor $Q$ of the fundamental oscillation mode was determined by ring-down measurements to be $Q_{\tn{inital}}=5.8\cdot 10^{5}$.

\begin{figure}[h]
\includegraphics [scale=0.8] {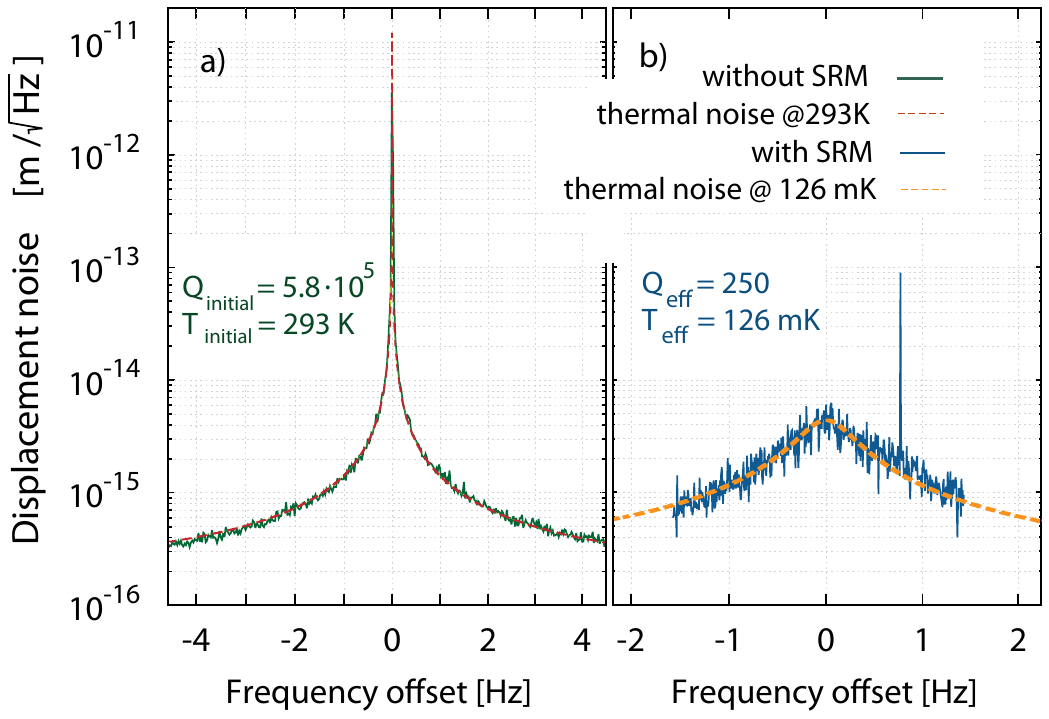}
\caption{Interferometer output spectra (a) without the SRM (\textit{i.e.} without the cavity) close to dark fringe, and (b) for a cavity-enhanced interferometer. The latter spectrum was taken at operation point $1$ [Fig.~\ref{outpower}(a)], with a detuning for which our theory predicts a particularly strong cooling effect. The input power was \unit{200}{\milli\watt} in (a) and (b).
The thermal noise levels (dashed lines) represent fits to the measurements with the mechanical $Q_{\tn{eff}}$ factor as the fitting parameter. In this measurement we could reach an effective mechanical quality factor $Q_{\tn{eff}}=\unit{250}{}$ and an effective temperature of $T_{\tn{eff}}=\unit{126}{\milli\kelvin}$.
}
\label{spectra}
\end{figure}

The generalized optomechanical coupling, which includes dispersive and dissipative coupling, was observed by detecting the interferometer's output power spectrum with a photodiode.
Fig.~\ref{spectra} shows two example spectra. The peaks correspond to the thermally excited motion of the membrane's fundamental resonance.
Both measurements were recorded when the membrane position was set such that the carrier light interfered almost destructively in the interferometer output port (position 1 in Fig.~\ref{outpower}).
The measurement in Fig.~\ref{spectra} (a) was performed without a SRM and calibrated using the method of Ref. \cite{Westphal2012}, whereas Fig.~\ref{spectra} (b) refers to a measurement with a (detuned) signal recycling cavity. The latter shows strong damping (optical cooling) of the membrane oscillation.
To derive the damped $Q$ factor we fitted the $Q_{\tn{eff}}$ value such that our model well described the height and the width of the thermally excited membrane resonance \cite{Yamamoto2010}. $Q_{\tn{eff}}$ here is the resulting $Q$ factor of the membrane which was influenced by the radiation pressure force (due to both dispersive and dissipative coupling). This adds an optically induced damping which changes the $Q$ factor, which is modeled in \cite{Tarabrin2013} and summarized in the Supplementary Material \cite{supplementary}.
\begin{figure}[h]
\includegraphics [scale=1.2] {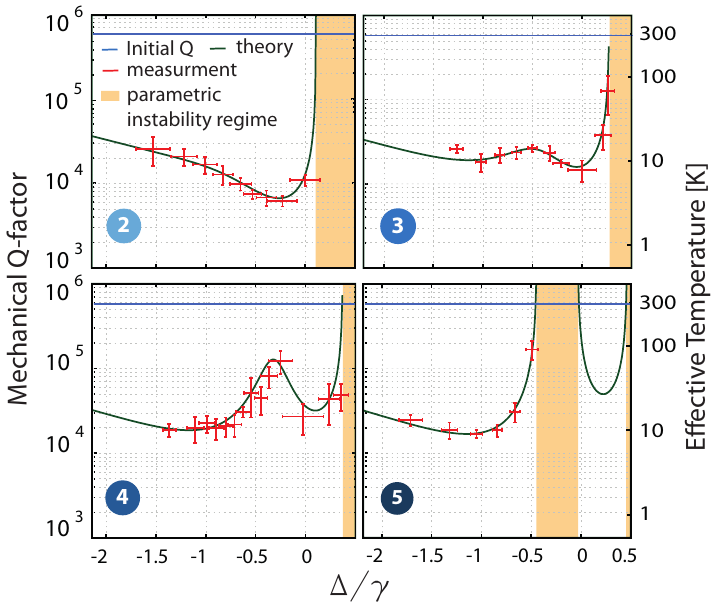}
\caption{Observed evidence of generalized optomechanical coupling and cooling on cavity resonance. We measured the effective mechanical $Q$ factor versus cavity detuning for four membrane positions as given in Fig.~\ref{outpower}(a) with an input power of $\unit{20}{\milli\watt}$. Already in the first graph cooling on cavity resonance is clearly visible. The next panels show an increasing influence of dissipative coupling and cooling regions that significantly expand into the region of positive (blue) detunings. The last panel confirms the theoretically predicted existence of a new instability region, which appears at negative (red) detunings. 
Solid lines refer to our theory of generalized optomechanical coupling. Regions of instability are marked yellow. The error bars represent the standard deviation of five independent measurements.
}
\label{cooling-dm}
\end{figure}

To observe generalized optomechanical coupling we exploited the tunability of dissipative and dispersive coupling in our setup, and we quantified the optical cooling at various positions of the membrane and SRM, see Fig.~\ref{cooling-dm}. The power inside the cavity was sufficiently low to not influence the measured quality factors by optical absorption of the membrane. In all measurements we positioned the membrane in the vicinity of a standing-wave node rather than in the vicinity of a standing-wave antinode \cite{Friedrich2011}. The frequency shift of the membrane resonance due to optical absorption was confirmed to be always less than \unit{500}{\hertz}. In this case the influence on the $Q$ factor is estimated to be on the order of a few percent and thus negligible for our analysis presented here.

Fig.~\ref{cooling-dm} represents the observation of generalized optomechanical coupling, i.e., strong signatures of interfering dispersive and dissipative couplings. All four graphs are distinct from conventional optomechanics with strongly dominating dispersive coupling. The first graph corresponds to membrane position $2$ in Fig.~\ref{outpower}. For negative detuning of the signal-recycling cavity we observed optical cooling ($Q_{\tn{eff}}<Q_{\tn{inital}}$), similar to the purely dispersive regime. But the same graph also shows optical cooling on cavity resonance (zero detuning). This effect is not possible in dispersive optomechanics. For larger positive detunings the membrane oscillation is parametrically heated ($Q_{\tn{eff}}>Q_{\tn{inital}}$). This effect is intrinsically unstable and eventually damped by the nonlinear behavior of the membrane for strong oscillations. Instability regions are marked as yellow areas.
In the second graph optical cooling on resonance is more pronounced, again in excellent agreement with our theory (solid line) \cite{Tarabrin2013, supplementary}.
The third graph shows a further evolution of the cooling spectrum. The optical cooling is observed up to detunings as large as $\Delta/\gamma=0.3$.
\begin{figure}[h]
\includegraphics [scale=1.15] {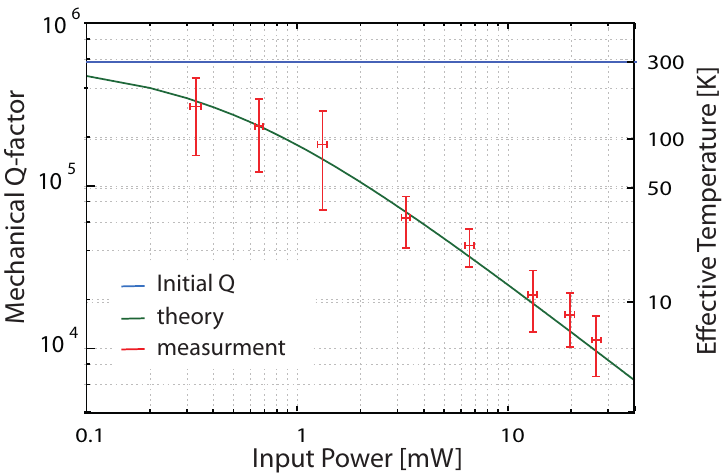}
\caption{Cooling on cavity resonance versus input power. The measurements are in excellent agreement with the theoretical model presented in Ref. \cite{Tarabrin2013,supplementary}. Here, the membrane was at position $3$, as described in Fig.~\ref{outpower}.
}
\label{coolingR}
\end{figure}
The last graph eventually shows the occurrence of a new instability region for small negative detunings. The newly appearing instability region is already visible as a reduced cooling performance in subfigures 3 and 4. Our theory predicts a well-separated cooling region at positive detunings which, however, could not be explored in the present experiment\footnote{The problem is that by passing through a region of instability the membrane is excited to very large amplitudes that cover the full range of displacements shown in Fig.~\ref{outpower}(a). Our linearized model and its predictions regarding (in)stabilities is not valid for such large amplitudes.}. The observed  behavior of the $Q$ factor is due to a complex structure of the radiation pressure noise spectral density: in the case of generalized optomechanical coupling this spectral density is a mixture of a Lorentz profile and a Fano profile corresponding to the dispersive and dissipative contributions respectively \cite{Elste2009,Xuereb2011,Tarabrin2013,supplementary}. Fig.~\ref{coolingR} shows optical cooling on resonance (measured at membrane position $3$) versus input power. Again, we find compelling agreement with the model summarized in the Supplementary Material \cite{supplementary}.

In conclusion, we realized a cavity-optomechanical setup with strong dissipative coupling between the cavity field and the mechanical oscillator, which did not rely on photon loss into unaccessible channels, in contrast to the experiments reported in Ref. \cite{Li2009,Wu2014}. In the language of the Hamiltonian description developed in Ref. \cite{Elste2009,Xuereb2011,Weiss2013} dispersive and dissipative coupling strengths can be characterized, respectively, by rates (per single photon) $g_\omega=x_\mathrm{ZPF}\,d\omega_\mathrm{c}(x)/dx|_{x_0}$ and $g_\gamma=x_\mathrm{ZPF}\,d\gamma(x)/dx|_{x_0}$, where $x_{\mathrm{ZPF}}=\sqrt{\hbar/(2m\omega_\mathrm{m})}$ is the amplitude of zero-point mechanical fluctuations of the membrane, $\omega_\mathrm{c}(x)$ and $\gamma(x)$ are the position-dependent cavity eigenfrequency and half-linewidth respectively, and $x_0$ is the mean (DC) position of the membrane. Explicit formulas for the coupling rates can be found in the Supplement Material \cite{supplementary}. For the parameters of the present setup these rates are tunable from 0 to about $0.1$ Hz, depending on the exact position of the membrane.

For the first time optical cooling of a mechanical oscillator through dissipative coupling, including, in particular, cooling on cavity resonance was observed. We measured a strong reduction in effective temperature of $3$ orders of magnitude. A reduction of the mechanical quality factor on resonance, as well as the existence of a second instability on the cooling side of the cavity resonance, are key predictions for a dissipative coupling in our experiment. We found excellent agreement with our model for generalized optomechanical coupling. Stronger cooling is predicted if the internal loss of the interferometer can be reduced. Ground state cooling is predicted outside the sideband-resolved regime \cite{Xuereb2011}. Our work might pave the way towards ground-state cooling of heavy objects. Since the dissipative coupling in our setup was external, the information gathered by photodiodes can in principle be used for  conditionally defining an almost pure mechanical quantum state as suggested in Refs. \cite{Mueller-Ebhardt2008,Mueller-Ebhardt2009} allowing for quantum physics with the motion of heavy objects.

Overall, we confirmed the theory of generalized optomechanical coupling in a regime that is of interest also in the field of gravitational wave detection. Dissipative coupling can give rise to a stable optical spring (that is a positive shift of the mechanical frequency and damping) \cite{Tarabrin2013}, as  proposed in Ref. \cite{Vostrosablin2014} for the improvement of GW detectors. The effects of a stable spring can in principle also be tested in our setup. The expected frequency increase due to an optical spring in our current setup, however, is small and masked by frequency changes due to absorption, i.e. absorptive heating.

~\\
\begin{acknowledgments}
This work was supported by the Marie Curie Initial Training Network cQOM,
by the ERC Advanced Grant MassQ, and by the International Max Planck Research School for Gravitational Wave Astronomy (IMPRS). H. Kaufer acknowledges support from the HALOSTAR scholarship program. K. Hammerer and S. Tarabrin acknowledge support through EU project iQUOEMS. The work of F. Khalili was supported by LIGO NSF grant  PHY-1305863and Russian Foundation for Basic Research grant No.11-02-00383-a.\end{acknowledgments}

~\\
~\\
\section{Appendix}
In this supplementary material we derive the formulas at the basis of the theory curves shown in the figures of the main text. Our treatment here essentially reproduces the derivations presented already in Ref.\,\cite{Tarabrin2013} (especially in Appendix A) with slight generalizations accommodating for the unbalanced central beamsplitter used in the present experiment.

\section{Propagation of fields}\label{sec_propag_fields}

\begin{figure}
\begin{center}
\includegraphics[scale=0.5]{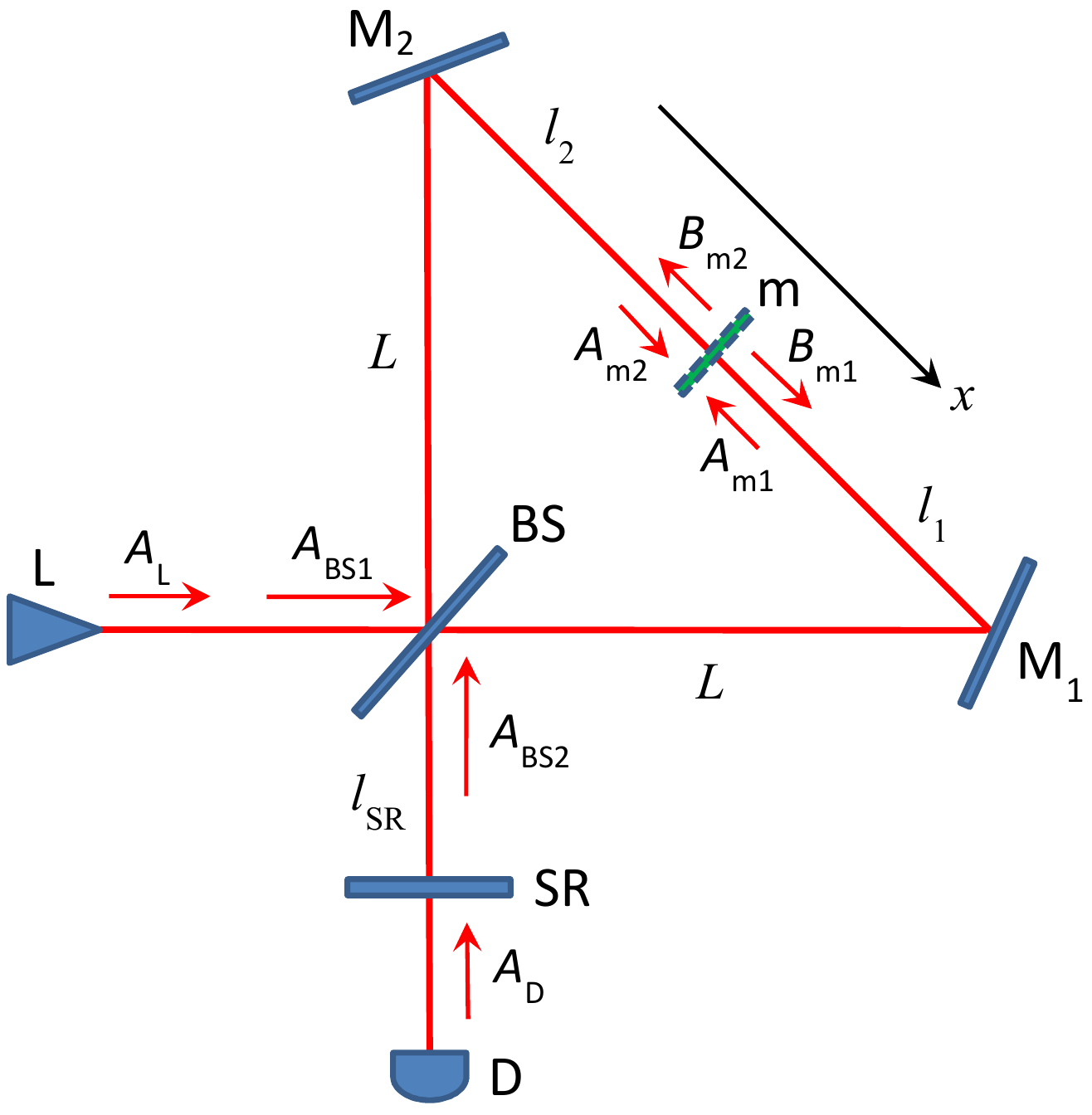}
\caption{Fields in a Michelson-Sagnac interferometer.}
\label{pic_MSR_interferometer}
\end{center}
\end{figure}

Consider a signal-recycled Michelson-Sagnac interferometer (MSI) as shown in Fig.\,\ref{pic_MSR_interferometer} with a central beamsplitter BS having amplitude reflectivity $r_\mathrm{BS}>0$ and transmissivity $t_\mathrm{BS}>0$, two steering mirrors $\textrm{M}_1$ and $\textrm{M}_2$ both having 100\% reflectivity, a semitransparent membrane m with amplitude reflectivity $r_\mathrm{m}>0$ and transmissivity $t_\mathrm{m}>0$, and a signal-recycling mirror SR with amplitude reflectivity $r_\mathrm{SR}>0$ and transmissivity $t_\mathrm{SR}>0$. The interferometer is driven by a laser L through the laser port. Photons emanating through the other, detector port impinge on a detector D. We denote the distance between SR mirror and BS as $l_\mathrm{SR}$, the arm length as $L$ and the distances between folding mirrors $\textrm{M}_1$ and $\textrm{M}_2$ and the membrane as $l_1 = l-\delta l/2$ and $l_2 = l+\delta l/2$, respectively. This means that $l_1+l_2=2l$, $l_2-l_1=\delta l$ and the position of the membrane on the $x$-axis is $x=\delta l/2$. The total mean length of the SR-m path is $\mathcal{L}=L+l+l_\mathrm{SR}$.

In any spatial location inside the interferometer we decompose the electric field of the coherent, plane and linearly polarized electromagnetic wave into the sum of a steady-state (mean) field with amplitude $A_0$ and carrier frequency $\omega_0$ (wavenumber $k_0=\omega_0/c$ and wavelength $\lambda_0=2\pi/k_0$), and slowly-varying (on the scale of $1/\omega_0$) perturbation field with amplitude $a(t)$ describing vacuum noises and the contribution from the motion of the membrane,
\begin{align*}
    A(t)&=\sqrt{\frac{2\pi\hbar\omega_0}{\mathcal{A}c}}\left[A_0e^{-i\omega_0t} + a(t)e^{-i\omega_0t}\right] + \textrm{{h.c.}},\\
    a(t)&=\int_{-\infty}^{+\infty}a(\omega_0+\Omega)e^{-i\Omega t}\,\frac{d\Omega}{2\pi}.
\end{align*}
Here $\mathcal{A}$ is the area of laser beam's cross-section and $c$ is the speed of light. Unless mentioned explicitly, we will deal with fields in the frequency domain only and omit frequency arguments for briefness.

The laser L emits a drive-wave $A_\mathrm{L}$ with mean amplitude $A_{\mathrm{L}0}$ and optical fluctuations $a_\mathrm{L}$. For simplicity we assume that there are no technical fluctuations so that the laser is shot-noise limited, $[a_\mathrm{L}(\omega_0+\Omega),a_\mathrm{L}^\dag(\omega_0+\Omega')]=2\pi\delta(\Omega-\Omega')$. The vacuum field $A_\mathrm{D}$ entering through the SR mirror (SRM) from detector port has zero mean amplitude but non-zero vacuum noise $a_\mathrm{D}$, uncorrelated with vacuum noise from the laser port and obeying the similar commutation relation $[a_\mathrm{D}(\omega_0+\Omega),a_\mathrm{D}^\dag(\omega_0+\Omega')]=2\pi\delta(\Omega-\Omega')$. We unite these into a vector-column of input fields $\textit{\textbf{A}}_\mathrm{in}=(A_\mathrm{L},A_\mathrm{D})$, so that the vector of mean input fields is $\textit{\textbf{A}}_{\mathrm{in}0}=(A_{\mathrm{L}0},0)$ and the vector of perturbation fields is $\textit{\textbf{a}}_\mathrm{in}=(a_\mathrm{L},a_\mathrm{D})$. Due to the linearity of the system input fields can be propagated throughout the interferometer as independent Fourier components.

Consider first the case without SRM and with a fixed membrane. The latter condition allows us to treat mean and perturbation fields on equal footing. Input fields (in this case coinciding with the fields incident on the beamsplitter) linearly transform into the output fields:
\begin{equation*}
    \textit{\textbf{A}}_{\mathrm{out}}=\mathbb{M}_\textrm{BS}^T\mathbb{P}_L\mathbb{P}_l\mathbb{M}_\textrm{m}\mathbb{P}_l\mathbb{P}_L
    \mathbb{M}_\textrm{BS}\textit{\textbf{A}}_{\mathrm{in}}\equiv\mathbb{M}_\textrm{MS}\textit{\textbf{A}}_{\mathrm{in}}.
\end{equation*}
Here
\begin{equation*}
    \mathbb{M}_\mathrm{BS}=
    \begin{pmatrix}
      t_\mathrm{BS} & -r_\mathrm{BS} \\
      r_\mathrm{BS} & t_\mathrm{BS} \\
    \end{pmatrix},\quad
    \mathbb{M}_\textrm{m}=
    \begin{pmatrix}
      r_\mathrm{m} & it_\mathrm{m} \\
      it_\mathrm{m} & r_\mathrm{m} \\
    \end{pmatrix},
\end{equation*}
are the transformation matrices of beamsplitter and membrane, each chosen in the form most convenient for calculations (as distinct from Ref.\,\cite{Tarabrin2013} where the  membrane matrix was chosen to be real), and
\begin{equation*}
    \mathbb{P}_L=
    \begin{pmatrix}
      e^{ikL} & 0 \\
      0 & e^{ikL} \\
    \end{pmatrix},\quad
    \mathbb{P}_l=
    \begin{pmatrix}
      e^{ikl_1} & 0 \\
      0 & e^{ikl_2} \\
    \end{pmatrix},
\end{equation*}
are the propagation matrices comprising the phase shifts along the horizontal/vertical arms (of length $L$) and diagonal half-arms (of lengths $l_{1,2}$). For mean fields one should apply the substitution $k=k_0$ and for perturbation fields $k=k_0+K=k_0+\Omega/c$. The matrix $\mathbb{M}_\textrm{MS}$ thus represents the transformation matrix of a non-recycled MSI,
\begin{equation*}
    \mathbb{M}_\textrm{MS}=e^{2ik(L+l)}
    \begin{pmatrix}
      \rho & i\tau \\
      i\tau & \rho^* \\
    \end{pmatrix},
\end{equation*}
with
\begin{subequations}
\begin{align}
    \rho&=r_\mathrm{m}\left(r_\mathrm{BS}^2e^{ik\delta l} + t_\mathrm{BS}^2e^{-ik\delta l}\right)+
    2it_\mathrm{m}r_\mathrm{BS}t_\mathrm{BS},\label{eq_MSI_reflectivity}\\
    \tau&=-ir_\mathrm{m}r_\mathrm{BS}t_\mathrm{BS}\left(e^{ik\delta l} - e^{-ik\delta l}\right)+
    t_\mathrm{m}(t_\mathrm{BS}^2-r_\mathrm{BS}^2).\label{eq_MSI_transmissivity}
\end{align}
\end{subequations}
Physically $\rho$ is the MSI amplitude reflectivity and $\tau$ is the amplitude transmissivity, such that $|\rho|^2+\tau^2=1$. Thus, a non-recycled MSI can be described as an effective mirror with reflectivity and transmissivity dependent on membrane position via $\delta l$ (cf. \cite{Tarabrin2013,Xuereb2011}). The dark port (dark fringe) condition for the interferometer corresponds to a choice for the membrane postion $\delta l$ such that $\tau=0$,
\begin{equation}
    \sin k_0\delta l=\frac{t_\mathrm{m}}{r_\mathrm{m}}\,\frac{r_\mathrm{BS}^2-t_\mathrm{BS}^2}{2r_\mathrm{BS}t_\mathrm{BS}}.
    \label{eq_dark_port_condition}
\end{equation}

If the SRM is inserted then the out-going field in the SR port is reflected back, such that the in-going fields incident on the beamsplitter are defined by the equation
\begin{equation}
    \textit{\textbf{A}}_\textrm{BS}=\mathbb{P}_\textrm{R}\mathbb{T}_\textrm{R}\textit{\textbf{A}}_{\textrm{in}}+
    \mathbb{P}_\textrm{R}\mathbb{R}_\textrm{R}\mathbb{P}_\textrm{R}\mathbb{M}_\textrm{MS}\textit{\textbf{A}}_\textrm{BS}.
    \label{eq_bs_fields_immov_membr}
\end{equation}
Here $\textit{\textbf{A}}_{\textrm{BS}}=(A_{\textrm{BS}1},A_{\textrm{BS}2})$ is the vector-column of in-going beamsplitter fields (see Fig.\,\ref{pic_MSR_interferometer}), $\mathbb{R}_\textrm{R}=\mathrm{diag}(0,r_\mathrm{SR})$ with zero standing for the absence of power-recycling mirror in laser port, $\mathbb{P}_\textrm{R}=\mathrm{diag}(1,e^{ikl_\textrm{SR}})$ is the propagation matrix in BS-SR path, and $\mathbb{T}_\textrm{R}=\mathrm{diag}(1,t_\mathrm{SR})$. Thus the first summand on the RHS of Eq.\,(\ref{eq_bs_fields_immov_membr}) stands for the input fields directly incident on the beamsplitter, while the second summand corresponds to a single round trip along the interferometer with reflection from the SRM. Solution of this equation yields
\begin{equation}
    \textit{\textbf{A}}_\textrm{BS}=(\mathbb{I}-\mathbb{P}_\textrm{R}\mathbb{R}_\textrm{R}\mathbb{P}_\textrm{R}\mathbb{M}_\textrm{MS})^{-1}\mathbb{P}_\textrm{R}\mathbb{T}_\textrm{R}
    \textit{\textbf{A}}_{\textrm{in}},
    \label{eq_bs_ingoing_fields_immov_membr}
\end{equation}
where $\mathbb{I}$ is the $2\times2$ unity matrix. We denote the inverse matrix in this solution as $\mathbb{K}_{\textrm{MSR}}$,
\begin{equation*}
    \mathbb{K}_{\textrm{MSR}}=\frac{1}{\mathcal{D}}
    \begin{pmatrix}
      \mathcal{D} & 0 \\
      r_\mathrm{SR}i\tau e^{2ik\mathcal{L}} & 1 \\
    \end{pmatrix},
\end{equation*}
This tells us that the MSI with SRM makes an effective Fabry-Perot cavity with associated resonance factor $1/\mathcal{D}$, where
\begin{equation}
    \mathcal{D}=1-r_\mathrm{SR}\rho^*e^{2ik\mathcal{L}}.
    \label{eq_resonance_factor}
\end{equation}

Note that the effective detuning of the laser carrier frequency from cavity resonance(s) is not solely defined by the corresponding shift in frequency (or cavity length) in contrast to the ordinary Fabry-Perot cavity. Denote $\arg\rho=\phi_{\textrm{DP}}+\delta\phi$, where $\phi_\mathrm{DP}=\arg\rho|_{\textrm{dark port}}$ is the phase of reflectivity of the MSI operated on dark port, and $\delta\phi$ is the deviation from it due to offset from dark port via membrane positioning and asymmetry of the beamsplitter. Assume that the following condition is satisfied on dark port: $2k_0\mathcal{L}-\phi_{\textrm{DP}}=2\pi N+2\delta_\mathrm{SR}\mathcal{L}/c$, where $N$ is fixed integer and $\delta_\mathrm{SR}\mathcal{L}/c\ll1$. This equation defines the detuning $\delta_\mathrm{SR}$ as the difference between the laser carrier frequency $\omega_0$ and the $N$-th resonance $\omega_\mathrm{res}$ of the MSI-SRM cavity when MSI is set on dark port:
\begin{equation*}
    \delta_\mathrm{SR}=\omega_0-\omega_\mathrm{res},\qquad\omega_\mathrm{res}=\frac{\pi Nc}{\mathcal{L}}+\frac{c\phi_{\textrm{DP}}}{2\mathcal{L}}.
\end{equation*}
This detuning can be generated either via tuning of the carrier frequency or via positioning of the SRM. In the latter case the detuning depends on the displacement $\delta\mathcal{L}$ of the SRM linearly, $\delta_\mathrm{SR}=-\omega_0\delta\mathcal{L}/\mathcal{L}$. However, when the MSI is operated off dark port ($\tau\neq0$), we need to introduce an additional detuning $\delta_\mathrm{MSI}$ associated with this offset via the equation $2k_0\mathcal{L}-\phi_{\textrm{DP}}-\delta\phi=2\pi N+2(\delta_\mathrm{SR}+\delta_\mathrm{MSI})\mathcal{L}/c$, from where it follows that $\delta_\mathrm{MSI}=-c\delta\phi/(2\mathcal{L})$. This detuning depends intricately on the position of the membrane and beamsplitter asymmetry $\varepsilon_\mathrm{BS}=r_\mathrm{BS}^2-t_\mathrm{BS}^2$ for the arbitrary offset from dark fringe, but for $k_0\delta l\ll1$ and $|\varepsilon_\mathrm{BS}|/\max(r_\mathrm{BS},t_\mathrm{BS})\ll1$ it can be calculated explicitly
\begin{equation*}
    \delta_\mathrm{MSI}=\frac{c}{2\mathcal{L}}\left[\mp r_\mathrm{m}t_\mathrm{m}\,\frac{(k_0\delta l)^2}{2}-r_\mathrm{m}^2
    \varepsilon_\mathrm{BS}k_0\delta l\pm r_\mathrm{m}t_\mathrm{m}\,\frac{\varepsilon_\mathrm{BS}^2}{2}\right],
\end{equation*}
where the sign depends on the choice of a particular dark fringe from Eq.\,(\ref{eq_dark_port_condition}). Note that in Ref.\,\cite{Tarabrin2013} this detuning was denoted as $\delta_\mathrm{m}$, since the beamsplitter was considered as perfectly balanced, so the offset form dark fringe could be generated solely via membrane positioning. Having defined the two contributions, the total detuning can now be introduced as
\begin{equation*}
    \Delta=\delta_\mathrm{SR}+\delta_\mathrm{MSI},
\end{equation*}
and the inverse resonance factor (for mean fields) now reads,
\begin{align*}
    \mathcal{D}&=1-r_\mathrm{SR}|\rho|e^{2ik_0\mathcal{L}-i\arg\rho}=1-r_\mathrm{SR}|\rho|e^{2i\Delta\mathcal{L}/c}.
\end{align*}

It follows from this equation that the half-linewidth of the MSI-SRM cavity, $\gamma=c(1-r_\mathrm{SR}|\rho|)/(2\mathcal{L})$,
in the narrow-band approximation, $1-r_\mathrm{SR}\approx t_\mathrm{SR}^2/2\ll1$ and $1-|\rho|\approx\tau^2/2\ll1$, also has two contributions:
\begin{equation}
    \gamma=\gamma_\mathrm{SR}+\gamma_\mathrm{MSI},\qquad
    \gamma_\mathrm{SR}=\frac{ct_\mathrm{SR}^2}{4\mathcal{L}},\qquad
    \gamma_\mathrm{MSI}=\frac{c\tau^2}{4\mathcal{L}}.
    \label{eq_linewidth}
\end{equation}
Therefore, the total cavity linewidth accounts for finite SRM transmittance and finite transmittance of the MSI operated off dark port; since $\tau=\tau(\delta l)$, the latter contribution describes modulation of the linewidth by the motion of the membrane, thus implementing dissipative coupling in the effective cavity, as discussed in \cite{Tarabrin2013,Xuereb2011}. Optical losses in the recycled interferometer can be described by adding an effective loss factor to the SRM transmittance, $t_\mathrm{SR}^2\rightarrow t_\mathrm{SR}^2+t_\mathrm{loss}^2$.

\section{Radiation pressure force}

The radiation pressure force exerted on the membrane can be determined through the fields on the membrane surfaces, see Fig.\,\ref{pic_MSR_interferometer},
\begin{equation}
    F(t)=-\frac{\mathcal{A}}{4\pi}\Bigl\langle A_{\textrm{m}1}^2(t)+B_{\textrm{m}1}^2(t)-A_{\textrm{m}2}^2(t)-B_{\textrm{m}2}^2(t)\Bigr\rangle,
    \label{eq_rad_pres_force}
\end{equation}
where averaging is performed over the period of electromagnetic oscillations. In-going fields on the beamsplitter defined by Eq.\,(\ref{eq_bs_ingoing_fields_immov_membr}) propagate along the arms and transform into the fields incident on the membrane $(A_{\textrm{m}1},A_{\textrm{m}2})=\textit{\textbf{A}}_\textrm{m}=\mathbb{P}_l\mathbb{P}_L\mathbb{M}_\textrm{BS}\textit{\textbf{A}}_\textrm{BS}$ and reflected from it $(B_{\textrm{m}1},B_{\textrm{m}2})=\textit{\textbf{B}}_\textrm{m}=\mathbb{M}_\textrm{m}\textit{\textbf{A}}_\textrm{m}$, see Fig.\,\ref{pic_MSR_interferometer}. In terms of input fields
\begin{subequations}
\begin{align}
    \textit{\textbf{A}}_\textrm{m}&=\mathbb{M}_\mathrm{inc}\textit{\textbf{A}}_\textrm{in},\
    \mathbb{M}_\mathrm{inc}=\mathbb{P}_l\mathbb{P}_L\mathbb{M}_\textrm{BS}\mathbb{K}_{\textrm{MSR}}\mathbb{P}_\textrm{R}\mathbb{T}_\textrm{R},
    \label{eq_membr_fields_inc_immov_membr}\\
    \textit{\textbf{B}}_\textrm{m}&=\mathbb{M}_\mathrm{ref}\textit{\textbf{A}}_\textrm{in},\
    \mathbb{M}_\mathrm{ref}=\mathbb{M}_\mathrm{m}\mathbb{P}_l\mathbb{P}_L\mathbb{M}_\textrm{BS}\mathbb{K}_{\textrm{MSR}}\mathbb{P}_\textrm{R}
    \mathbb{T}_\textrm{R}.
    \label{eq_membr_fields_ref_immov_membr}
\end{align}
\end{subequations}
The components of matrix $\mathbb{M}_\mathrm{inc}$ are
\begin{widetext}
\begin{align*}
    \mathbb{M}_\mathrm{inc}^{(1,1)}&=\mathcal{D}^{-1}\Bigl[t_\mathrm{BS}\left(1-r_\mathrm{m}r_\mathrm{SR}e^{2ik(\mathcal{L}+\delta l/2)}\right)
    +r_\mathrm{BS}it_\mathrm{m}r_\mathrm{SR}e^{2ik\mathcal{L}}\Bigr]e^{ik(L+l-\delta l/2)},\\
    \mathbb{M}_\mathrm{inc}^{(1,2)}&=-\mathcal{D}^{-1}t_\mathrm{SR}r_\mathrm{BS}e^{ik(\mathcal{L}-\delta l/2)},\\
    \mathbb{M}_\mathrm{inc}^{(2,1)}&=\mathcal{D}^{-1}\Bigl[r_\mathrm{BS}\left(1-r_\mathrm{m}r_\mathrm{SR}e^{2ik(\mathcal{L}-\delta l/2)}\right)
    +t_\mathrm{BS}it_\mathrm{m}r_\mathrm{SR}e^{2ik\mathcal{L}}\Bigr]e^{ik(L+l+\delta l/2)},\\
    \mathbb{M}_\mathrm{inc}^{(2,2)}&=\mathcal{D}^{-1}t_\mathrm{SR}t_\mathrm{BS}e^{ik(\mathcal{L}+\delta l/2)},
\end{align*}
and of matrix $\mathbb{M}_\mathrm{ref}$
\begin{align*}
    \mathbb{M}_\mathrm{ref}^{(1,1)}&=\mathcal{D}^{-1}\Bigl[
    t_\mathrm{BS}\left(r_\mathrm{m}-r_\mathrm{SR}e^{2ik(\mathcal{L}+\delta l/2)}\right)+it_\mathrm{m}r_\mathrm{BS}e^{ik\delta l}\Bigr]e^{ik(L+l-\delta l/2)},\\
    \mathbb{M}_\mathrm{ref}^{(1,2)}&=\mathcal{D}^{-1}t_\mathrm{SR}\left(-r_\mathrm{BS}r_\mathrm{m}+it_\mathrm{m}t_\mathrm{BS}e^{ik\delta l}\right)e^{ik(\mathcal{L}-\delta l/2)},\\
    \mathbb{M}_\mathrm{ref}^{(2,1)}&=\mathcal{D}^{-1}\Bigl[
    r_\mathrm{BS}\left(r_\mathrm{m}-r_\mathrm{SR}e^{2ik(\mathcal{L}-\delta l/2)}\right)+it_\mathrm{m}t_\mathrm{BS}e^{-ik\delta l}\Bigr]e^{ik(L+l+\delta l/2)},\\
    \mathbb{M}_\mathrm{ref}^{(2,2)}&=\mathcal{D}^{-1}t_\mathrm{SR}\left(t_\mathrm{BS}r_\mathrm{m}-it_\mathrm{m}r_\mathrm{BS}e^{-ik\delta l}\right)e^{ik(\mathcal{L}+\delta l/2)}.
\end{align*}
\end{widetext}
Since the transfer matrices depend on frequency, we denote them as $\mathbb{M}_\mathrm{inc}(0)=\mathbb{M}_\mathrm{inc}|_{k=k_0}$, $\mathbb{M}_\mathrm{ref}(0)=\mathbb{M}_\mathrm{ref}|_{k=k_0}$ for mean fields at the laser frequency $\omega_0$, and $\mathbb{M}_\mathrm{inc}(\Omega)=\mathbb{M}_\mathrm{inc}|_{k=k_0+K}$, $\mathbb{M}_\mathrm{ref}(\Omega)=\mathbb{M}_\mathrm{ref}|_{k=k_0+K}$ for perturbation fields at sideband frequency $\Omega$.

To calculate the dynamical back-action we now need to take into account the motion of the membrane. Consider the position operator $x_\mathrm{m}(t)$ with a corresponding Fourier-transformed operator $x_\mathrm{m}(\Omega)$. According to perturbation theory the fields on the membrane surfaces will have contributions of zeroth and first order in the mechanical displacement. One finds
\begin{align*}
    \textit{\textbf{B}}_{\textrm{m}0}&=\mathbb{M}_\textrm{m}\textit{\textbf{A}}_{\textrm{m}0},\\
    \textit{\textbf{b}}_{\textrm{m}}&= \mathbb{M}_\textrm{m}\textit{\textbf{a}}_{\textrm{m}}-2ik_0x_\textrm{m}r_\mathrm{m}\sigma_3\textit{\textbf{A}}_{\textrm{m}0},
\end{align*}
where $\sigma_3=\mathrm{diag}(1,-1)$. Thus the perturbation fields now contain both optical noises and the displacement of the membrane. Since the treatment of mean fields remains unchanged, we consider only the perturbation terms. The in-going fields on the beamsplitter are defined by the equation
\begin{multline*}
    \textit{\textbf{a}}_\textrm{BS}=\mathbb{P}_\textrm{R}\mathbb{T}_\textrm{R}\textit{\textbf{a}}_\textrm{in}+
    \mathbb{P}_\textrm{R}\mathbb{R}_\textrm{R}\mathbb{P}_\textrm{R}\mathbb{M}_\textrm{MS}\textit{\textbf{a}}_{\textrm{BS}}\\
    -\mathbb{P}_\textrm{R}\mathbb{R}_\textrm{R}\mathbb{P}_\textrm{R}\mathbb{M}_\textrm{BS}^T\mathbb{P}_L\mathbb{P}_l\,
    2ik_0x_\textrm{m}(\Omega)r_\mathrm{m}\sigma_3\textit{\textbf{A}}_{\textrm{m}0},
\end{multline*}
with solution
\begin{multline*}
    \textit{\textbf{a}}_\textrm{BS}=\mathbb{K}_{\textrm{MSR}}\mathbb{P}_\textrm{R}\mathbb{T}_\textrm{R}\textit{\textbf{a}}_{\textrm{in}}\\
    -2ik_0x_\textrm{m}\mathbb{K}_{\textrm{MSR}}\mathbb{P}_\textrm{R}\mathbb{R}_\textrm{R}\mathbb{P}_\textrm{R}\mathbb{M}_\textrm{BS}^T
    \mathbb{P}_L\mathbb{P}_lr_\mathrm{m}\sigma_3\textit{\textbf{A}}_{\textrm{m}0}.
\end{multline*}
Thus the incident and reflected fields on the membrane surfaces are
\begin{subequations}
\begin{align}
    \textit{\textbf{a}}_\textrm{m}&=
    \mathbb{M}_\mathrm{inc}(\Omega)\textit{\textbf{a}}_{\textrm{in}}-2ik_0x_\textrm{m}\mathbb{M}_x\sigma_3\textit{\textbf{A}}_{\textrm{m}0},
    \label{eq_membr_fields_inc_mov_membr}\\
    \textit{\textbf{b}}_\textrm{m}&=
    \mathbb{M}_\mathrm{ref}(\Omega)\textit{\textbf{a}}_{\textrm{in}}-2ik_0x_\textrm{m}(r_\mathrm{m}\mathbb{I}+
    \mathbb{M}_\textrm{m}\mathbb{M}_{x})\sigma_3\textit{\textbf{A}}_{\textrm{m}0}.
    \label{eq_membr_fields_ref_mov_membr}
\end{align}
\end{subequations}
The components of the matrix
\begin{equation*}
    \mathbb{M}_{x}=\mathbb{P}_l\mathbb{P}_L\mathbb{M}_\textrm{BS}\mathbb{K}_{\textrm{MSR}}\mathbb{P}_\textrm{R}\mathbb{R}_\textrm{R}\mathbb{P}_\textrm{R}
    \mathbb{M}_\textrm{BS}^T\mathbb{P}_L\mathbb{P}_lr_\mathrm{m}.
\end{equation*}
are
\begin{align*}
    \mathbb{M}_{x}^{(1,1)}&=\mathcal{D}^{-1}r_\mathrm{m}r_\mathrm{BS}^2r_\mathrm{SR}e^{2ik(\mathcal{L}-\delta l/2)},\\
    \mathbb{M}_{x}^{(2,2)}&=\mathcal{D}^{-1}r_\mathrm{m}t_\mathrm{BS}^2r_\mathrm{SR}e^{2ik(\mathcal{L}+\delta l/2)},\\
    \mathbb{M}_{x}^{(1,2)}&=\mathbb{M}_{ax}^{(2,1)}=-\mathcal{D}^{-1}r_\mathrm{m}r_\mathrm{BS}t_\mathrm{BS}r_\mathrm{SR}e^{2ik\mathcal{L}}.
\end{align*}

Substituting mean fields (\ref{eq_membr_fields_inc_immov_membr},\,\ref{eq_membr_fields_ref_immov_membr}) and perturbations fields (\ref{eq_membr_fields_inc_mov_membr},\,\ref{eq_membr_fields_ref_mov_membr}) into Eq.\,(\ref{eq_rad_pres_force}), ignoring the D.C. part and linearizing with respect to perturbation terms, one ends up with $F(\Omega)=F_\mathrm{BA}(\Omega)+F_x(\Omega)$. Here $F_\mathrm{BA}$ is the radiation pressure noise:
\begin{align*}
    F_{\textrm{BA}}(\Omega)=&-2\hbar k_0r_\mathrm{m}\textit{\textbf{A}}_{\textrm{in}0}^{*\,T}\mathbb{M}_\mathrm{inc}^{*\,T}(0)\sigma_3
    \mathbb{M}_\mathrm{ref}(\Omega)\textit{\textbf{a}}_\textrm{in}(\omega_0+\Omega)\nonumber\\
    &-2\hbar k_0r_\mathrm{m}\textit{\textbf{A}}_{\textrm{in}0}^{T}\mathbb{M}_\mathrm{inc}^T(0)\sigma_3
    \mathbb{M}_\mathrm{ref}^*(-\Omega)\textit{\textbf{a}}_\textrm{in}^\dag(\omega_0-\Omega),
\end{align*}
and $F_x(\Omega)=-\mathcal{K}(\Omega)x_\mathrm{m}(\Omega)$ is the ponderomotive force, i.e. dynamical part of the radiation pressure force caused by the motion of the membrane. The coefficient $\mathcal{K}(\Omega)$ modifies the dynamics of the membrane, and therefore represents the dynamical back-action,
\begin{align*}
    \mathcal{K}(\Omega)&=-\frac{2ik_0}{c}\,r_\mathrm{m}P_\textrm{in}\left[\mathbb{K}_{(1,1)}(\Omega)-\mathbb{K}_{(1,1)}^*(-\Omega)\right],\\
    \mathbb{K}(\Omega)&=\mathbb{M}_\mathrm{inc}^{*\,T}(0)\sigma_3\mathbb{M}_\mathrm{m}\Bigl[\mathbb{I}+2\mathbb{M}_{x}(\Omega)\Bigr]
    \sigma_3\mathbb{M}_\mathrm{inc}(0).
\end{align*}

The corresponding time-domain equation of motion of the membrane reads
\begin{equation*}
    \ddot{x}_\mathrm{m}+2\gamma_\mathrm{m}\dot{x}_\mathrm{m}+\omega_\mathrm{m}^2x_\mathrm{m}=
    \frac{F_\mathrm{BA}(t)}{m}+\frac{F_x(x_\mathrm{m},t)}{m}+\frac{G(t)}{m},
\end{equation*}
where $\gamma_\mathrm{m}$ is the intrinsic mechanical damping rate, $\omega_\mathrm{m}$ is the intrinsic mechanical frequency, $m$ is the membrane's effective mass, and $G(t)$ is the sum of all other external forces exerted on the membrane. Denote $K(\Omega)\equiv\Re[\mathcal{K}(\Omega)]$ and $\Gamma(\Omega)\equiv-\frac{1}{2}\Im[\mathcal{K}(\Omega)]/\Omega$. Then in frequency domain the equation of motion transforms to
\begin{multline}
    x_\mathrm{m}(\Omega)\Biggl[-\Omega^2-2i\left(\gamma_\mathrm{m}+\frac{\Gamma(\Omega)}{m}\right)\Omega+
    \left(\omega_\mathrm{m}^2+\frac{K(\Omega)}{m}\right)\Biggr]\\
    =\frac{1}{m}\,F_\textrm{BA}(\Omega)+\frac{1}{m}\,G(\Omega).
\end{multline}
In this equation $\Gamma/m$ modifies the damping rate (optical damping), and $K/m$ shifts the square of the mechanical frequency (optical spring). Both quantities are functions of $\Omega$ and also depend parametrically on the detuning $\Delta$ of laser carrier frequency from MSI-SRM cavity resonance. For a high quality oscillator one can neglect the frequency dependence, and approximate $\Omega=\omega_\mathrm{m}$, such that the optical spring and damping can be considered as the functions of detuning only, $\mathcal{K}(\Omega,\Delta)=\mathcal{K}(\omega_\mathrm{m},\Delta)$. Since the shift of the mechanical frequency due to optical spring effect is small (relative to $\omega_\mathrm{m}$), we can define the effective mechanical quality factor modified by optical damping as
\begin{equation}
    Q(\Delta)=\frac{1}{2}\,\frac{\omega_\mathrm{m}}{\gamma_\mathrm{m}+\Gamma(\omega_\mathrm{m},\Delta)/m}.
\end{equation}
This equation is used to plot the theoretical curves in Fig.\,4 of the main text.

\section{Hamiltonian description}
\begin{figure}
\begin{center}
\includegraphics[scale=0.75]{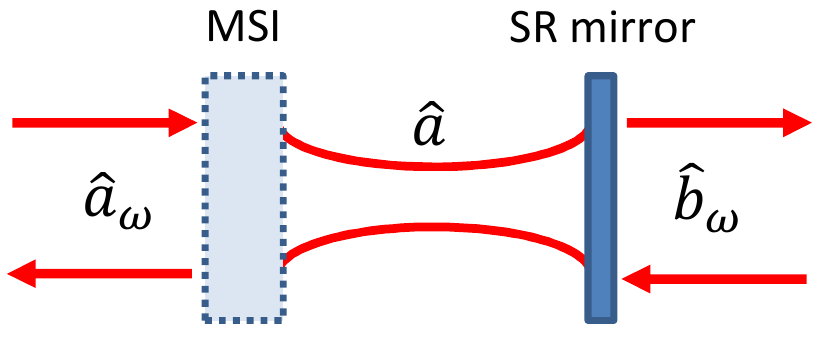}
\caption{Schematic of the effective MSI-SRM cavity.}
\label{pic_cavity}
\end{center}
\end{figure}

Although the above derivations have been performed using the transfer matrix approach, our setup also admits a description in terms of a Hamiltonian for a single effective cavity mode, in analogy to Ref.\,\cite{Xuereb2011}. The Hamiltonian of the single optical mode in the effective MSI-SRM cavity interacting with external fields and the membrane reads (see Fig.\,\ref{pic_cavity}),
\begin{align}
    \hat{H}={}&\omega_\mathrm{c}(\hat{x})\hat{a}^\dag\hat{a}+\omega_\mathrm{m}\hat{c}^\dag\hat{c}+
    \int\omega\left(\hat{a}_\omega^\dag\hat{a}_\omega+\hat{b}_\omega^\dag\hat{b}_\omega\right)d\omega\nonumber\\
    &+i\sqrt{2\gamma_\mathrm{MSI}(\hat{x})}\int\left(\hat{a}_\omega^\dag\hat{a}-\hat{a}_\omega\hat{a}^\dag\right)\frac{d\omega}{\sqrt{2\pi}}\nonumber\\
    &+i\sqrt{2\gamma_\mathrm{SR}}\int\left(\hat{b}_\omega^\dag\hat{a}-\hat{b}_\omega\hat{a}^\dag\right)\frac{d\omega}{\sqrt{2\pi}}.
    \label{eq_general_Hamiltonian}
\end{align}
Here $\omega_\mathrm{c}(\hat{x})$ is the eigenfrequency of the intracavity mode dependent on the position $\hat{x}$ of the membrane, $\hat{a}$ is the intracavity mode annihilation operator obeying the commutation relation $[\hat{a},\hat{a}^\dag]=1$, $\hat{a}_\omega$ and $\hat{b}_\omega$ are the annihilation operators of the external fields coupled through MSI and SRM respectively, obeying $[\hat{a}_\omega,\hat{a}_\omega^\dag]=[\hat{b}_\omega,\hat{b}_\omega^\dag]=\delta(\omega-\omega')$, $\omega_\mathrm{m}$ is the intrinsic mechanical frequency of the membrane, $\hat{c}$ is the annihilation operator of the mechanical motion, such that $\hat{x}=x_\mathrm{ZPF}(\hat{c}+\hat{c}^\dag)/\sqrt{2}$ with $x_{\mathrm{ZPF}}=\sqrt{\hbar/(2m\omega_\mathrm{m})}$ being the amplitude of zero-point mechanical fluctuations. Half-linewidths $\gamma_\mathrm{SR}$ and $\gamma_\mathrm{MSI}$ are defined in Eq.\,(\ref{eq_linewidth}). The above Hamiltonian generalizes the one in Ref.\,\cite{Xuereb2011} for the finite-transmittive SRM. Note however, that SRM with finite transmittance makes the effective cavity a double-sided one, therefore at least a pair of dissipative coupling strengths should be included in the Hamiltonian. But since our setup is not driven by a laser through detector port, second dissipative coupling channel can be ignored.

Usually one expands $x$-dependent parameters of the cavity to first order near some mean position $x_0$ of the oscillator to arrive at the linear-coupling model: $\omega_\mathrm{c}(x)\approx\omega_\mathrm{c}(x_0)+g_\omega\xi$ and $\gamma_\mathrm{MSI}(x)\approx\gamma_\mathrm{MSI}(x_0)+g_\gamma\xi$, where $\xi=(x-x_0)/x_\mathrm{ZPF}$ is the normalized displacement of the mechanical oscillator, $g_\omega=d\omega_\mathrm{c}/dx|_{x_0}$ is the dispersive coupling constant, and $g_\gamma=d\gamma_\mathrm{MSI}/dx|_{x_0}$ is the dissipative coupling constant. Then the interaction Hamiltonian in Eq.\,(\ref{eq_general_Hamiltonian}) takes the form,
\begin{equation*}
    \hat{H}_\mathrm{int}=g_\omega\hat{\xi}\hat{a}^\dag\hat{a}+
    i\,\frac{g_\gamma}{\sqrt{2\gamma_\mathrm{MSI}}}\,\hat{\xi}
    \int\left(\hat{a}_\omega^\dag\hat{a}-\hat{a}_\omega\hat{a}^\dag\right)\frac{d\omega}{\sqrt{2\pi}}.
\end{equation*}

The coupling constants $g_\omega$ and $g_\gamma$ can be derived using their definitions. According to Eq.\,(\ref{eq_resonance_factor}), the eigenfrequency of the intracavity mode is defined by the equation $2\omega\mathcal{L}/c-\arg\rho=2\pi N$ with fixed integer $N$,
\begin{equation*}
    \omega_\mathrm{c}(x)=\frac{\pi Nc}{\mathcal{L}}+\frac{c}{2\mathcal{L}}\,\arg\rho(x).
\end{equation*}
It follows that $\omega_\mathrm{c}'(x)=c\arg'\rho(x)/(2\mathcal{L})$. Simple differentiation leads to $\arg'\rho(x)=\cos^2(\arg\rho)\tan'(\arg\rho)$. Substituting here $\cos(\arg\rho)=\mathfrak{R}[\rho]/|\rho|$, $\tan(\arg\rho)=\mathfrak{I}[\rho]/\mathfrak{R}[\rho]$, and rewriting Eqs.\,(\ref{eq_MSI_reflectivity},\ref{eq_MSI_transmissivity}) in terms of the mean membrane position $x=\delta l/2$ and beamslitter asymmetry $\varepsilon_\mathrm{BS}=r_\mathrm{BS}^2-t_\mathrm{BS}^2$,
\begin{align*}
    \rho(x)=&r_\mathrm{m}\cos2k_0x+i\left(r_\mathrm{m}\varepsilon_\mathrm{BS}\sin2k_0x+t_\mathrm{m}\sqrt{1-\varepsilon_\mathrm{BS}^2}\right),\\
    \tau(x)=&r_\mathrm{m}\sqrt{1-\varepsilon_\mathrm{BS}^2}\sin2k_0x-t_\mathrm{m}\varepsilon_\mathrm{BS},
\end{align*}
one finally arrives at,
\begin{equation*}
    g_\omega=\omega_0\,\frac{x_\mathrm{ZPF}}{\mathcal{L}}\,
    \frac{r_\mathrm{m}^2\varepsilon_\mathrm{BS}+r_\mathrm{m}t_\mathrm{m}\sqrt{1-\varepsilon_\mathrm{BS}^2}\sin2k_0x_0}{|\rho(x_0)|^2}.
\end{equation*}

Similarly, one derives $g_\gamma$ using the definition of $\gamma_\mathrm{MSI}$ in Eq.\,(\ref{eq_linewidth}). Simple calculation of $\gamma_\mathrm{MSI}'(x)=c\tau(x)\tau'(x)/(2\mathcal{L})$ leads to:
\begin{equation*}
    g_\gamma=\omega_0\,\frac{x_\mathrm{ZPF}}{\mathcal{L}}\,\tau(x_0)r_\mathrm{m}\sqrt{1-\varepsilon_\mathrm{BS}^2}\cos2k_0x_0.
\end{equation*}
One must be careful when evaluating $g_\gamma/\sqrt{2\gamma_\mathrm{MSI}}$ at the dark port, since both the numerator and denominator vanish at this operation point. However, their ratio, proportional to $\tau'(x_0)\sim\cos2k_0x_0$, remains finite.

\begin{figure}[H]
\begin{center}
\includegraphics[scale=0.42]{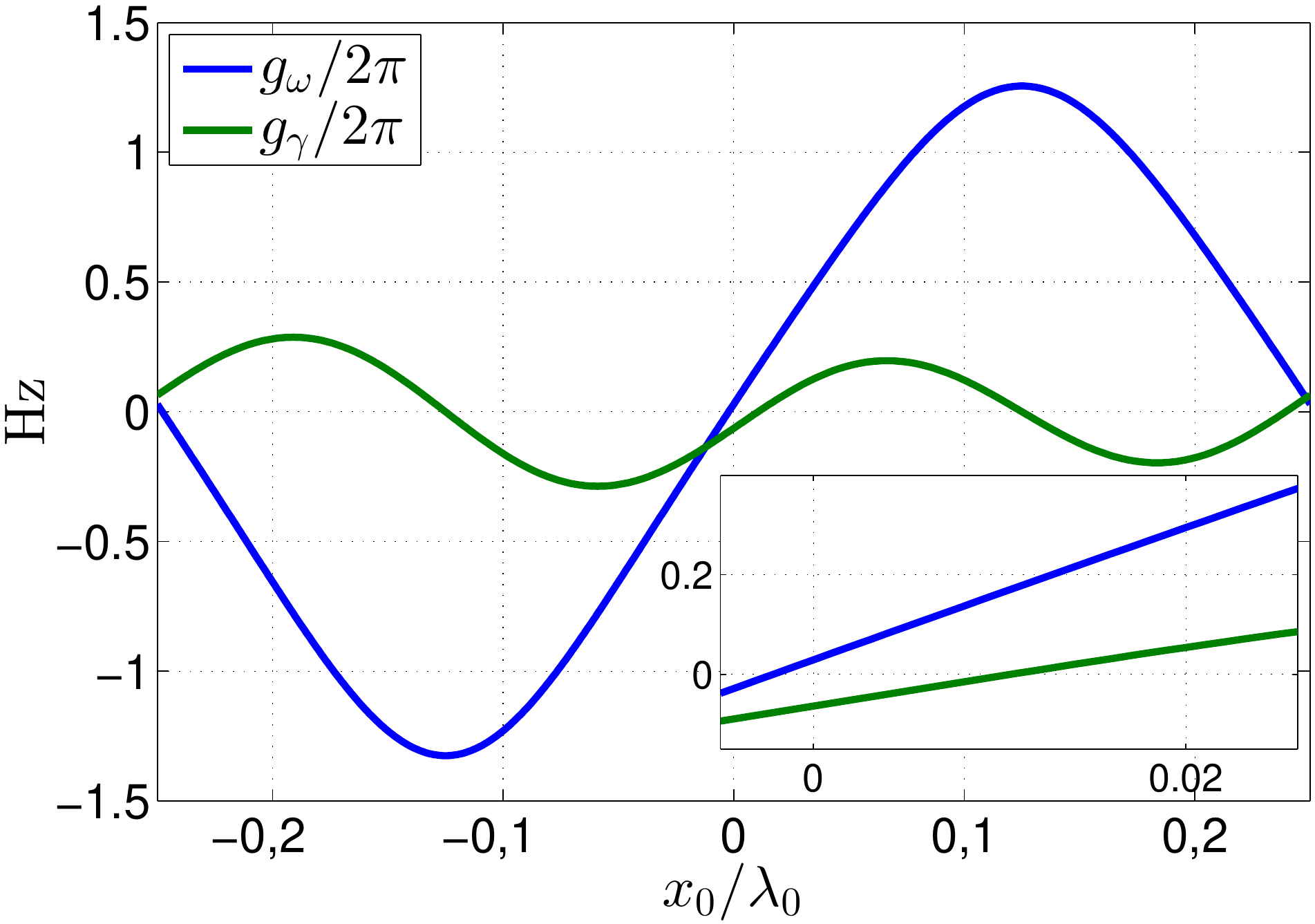}
\caption{Coupling constants versus mean position of the membrane with the experimentally relevant parameters: $\lambda_0=1064$ nm, $\mathcal{L}=8.7$ cm, $r_\mathrm{m}^2=0.17$, $\varepsilon_\mathrm{BS}=0.06$, $m=80$ ng, $\omega_\mathrm{m}/2\pi=136$ kHz.}
\label{pic_coupling_constants}
\end{center}
\end{figure}

Fig.\,\ref{pic_coupling_constants} shows the coupling constants plotted versus the mean position of the membrane for the experimentally relevant parameters specified in the main text. The inset shows the coupling constants in the experimentally relevant limits for $x_0/\lambda_0$.\\

\section{Parameters of the experimental setup} 
\begin{table}[h]
    \begin{tabular}{l l l l}
        \centering
        \textbf{Parameter} & \textbf{Notation} & \textbf{Value} \\
        \hline
        Effective cavity length & $\mathcal{L}$ & 8.7 cm\\
        Laser carrier wavelength & $\lambda_0$ & 1064 nm\\
        Laser input power & $P_\mathrm{in}$ & $0.3-200$ mW\\
        Membrane power reflectivity & $r_\mathrm{m}^2$ & 0.17\\
        Beamsplitter asymmetry & $r_\mathrm{BS}^2-t_\mathrm{BS}^2$ & $6\cdot10^{-2}$\\
        SRM power reflectivity & $r_\mathrm{SR}^2$ & 0.9997\\
        Optical losses inside the interferometer& $t_\mathrm{loss}^2$ & $5\cdot10^{-3}$\\
        Membrane mechanical frequency & $\omega_\mathrm{m}/{2\pi}$ & 136 kHz\\
        Membrane intrinsic $Q$-factor & $Q_\mathrm{m}$ & $5.8\cdot10^5$\\
        Membrane effective mass & $m$ & 80 ng\\
    \end{tabular}
\end{table}
%

\end{document}